\documentclass[11pt]{article}
\usepackage{Blois_modCEdip,epsfig,wick}

\def\be{\begin{equation}}
\def\ee{\end{equation}}
\def\bea{\begin{eqnarray}}
\def\eea{\end{eqnarray}}

\renewcommand\slash[1]{\not \! #1}

\newcommand{\rightslash}{\! \stackrel{\rightarrow}{\slash{\partial}}}
\newcommand{\leftslash}{\! \stackrel{\leftarrow}{\slash{\partial}}}

\begin{document}



\vspace*{4cm}
\title{ON THE NONPERTURBATIVE FOUNDATIONS OF THE DIPOLE PICTURE}

\author{\underline{C.\ EWERZ}$^{a,b}$, O.\ NACHTMANN$^c$}

\address{$^a$ Dipartimento di Fisica, Universit{\`a} di Milano and INFN, Sezione di Milano\\
Via Celoria 16, I-20133 Milano, Italy\\
$^b$ ECT*, Strada delle Tabarelle 286, I-38050 Villazzano (Trento), Italy\\
$^c$ Institut f\"ur Theoretische Physik, Universit\"at Heidelberg\\
Philosophenweg 16, D-69120 Heidelberg, Germany}

\maketitle\abstracts{
Starting from a genuinely nonperturbative formulation of
photon-proton scattering we discuss which approximations
and assumptions are required to obtain the dipole picture
of high energy scattering. 
}

\section{Introduction}
\label{sec:intro}

In the last few years the dipole picture of deep inelastic photon-hadron 
scattering~\cite{Nikolaev:1990ja,Mueller:1993rr} 
has been widely used as a framework for interpreting HERA data, 
in particular in view of potential saturation effects at high energies. 
In the dipole picture the photon splits into a quark-antiquark pair 
-- a colour dipole -- which subsequently scatters off the proton. 
Accordingly, the cross section for transversely ($T$) or longitudinally 
($L$) polarised photons is given by 
\be
\label{dipolecross}
\sigma_{T,L} (x, Q^2) =  \sum_q \int d^2R_T \int_0^1 d\alpha 
\left| \psi^{(q)}_{T,L} (\alpha,\mathbf{R}_T,Q) \right|^2
\sigma^{(q)}_{\rm red} (R_T^2,s) \,,
\ee
where $\sqrt{s}$ is the energy and $x=Q^2/s$ is the Bjorken scaling 
variable with the photon virtuality $Q^2$. 
The photon wave function $\psi_{T,L}^{(q)}$ describes in leading 
order in perturbation theory the splitting 
of the photon into a quark and an antiquark of flavour $q$ with relative 
separation $\mathbf{R}_T$ in transverse space, carrying the 
longitudinal momentum fractions $\alpha$ and $1-\alpha$, respectively. 
The reduced cross section $\sigma_{\rm red}$ describes the dipole-proton 
scattering. 
Note that according to Eq.\ (\ref{dipolecross}) the energy dependence of the 
cross section is contained only in $\sigma_{\rm red}$. 
In practical applications of the dipole picture one fits the data by 
a suitable choice of the reduced cross section $\sigma_{\rm red}$. 
A prominent example is the Golec-Biernat-W\"usthoff 
model~\cite{Golec-Biernat:1998js} for $\sigma_{\rm red}$ which 
incorporates saturation effects. 

The dipole picture in the form described above is perturbatively motivated 
and is thus expected to be valid for large photon virtualities and small values of 
$x$. Since at HERA small $x$ is correlated with smaller virtualities, however, 
the actual interest is mostly concentrated on the region of medium 
(and sometimes even small) virtualities where potential saturation effects 
could be observed. It is therefore an important question whether the simple 
dipole picture as given by the above formula is complete, or whether 
there are contributions to photon-proton scattering that cannot be 
accommodated by Eq.\ (\ref{dipolecross}). More generally, it is crucial to 
understand which assumptions and approximations are required in order  
to obtain the dipole picture starting from a completely nonperturbative 
framework, and which potential corrections consequently need to be 
considered. 

In the following we report in condensed form on a recent 
study~\cite{Ewerz:2004vf,Ewerz:2005dipII} in which we address these 
questions. For a detailed presentation we refer the interested reader to 
these references. 

\section{Functional Methods for the Compton Amplitude}
\label{sec:functmeth}

\begin{figure}
\begin{center}
\includegraphics[width=14.45cm]{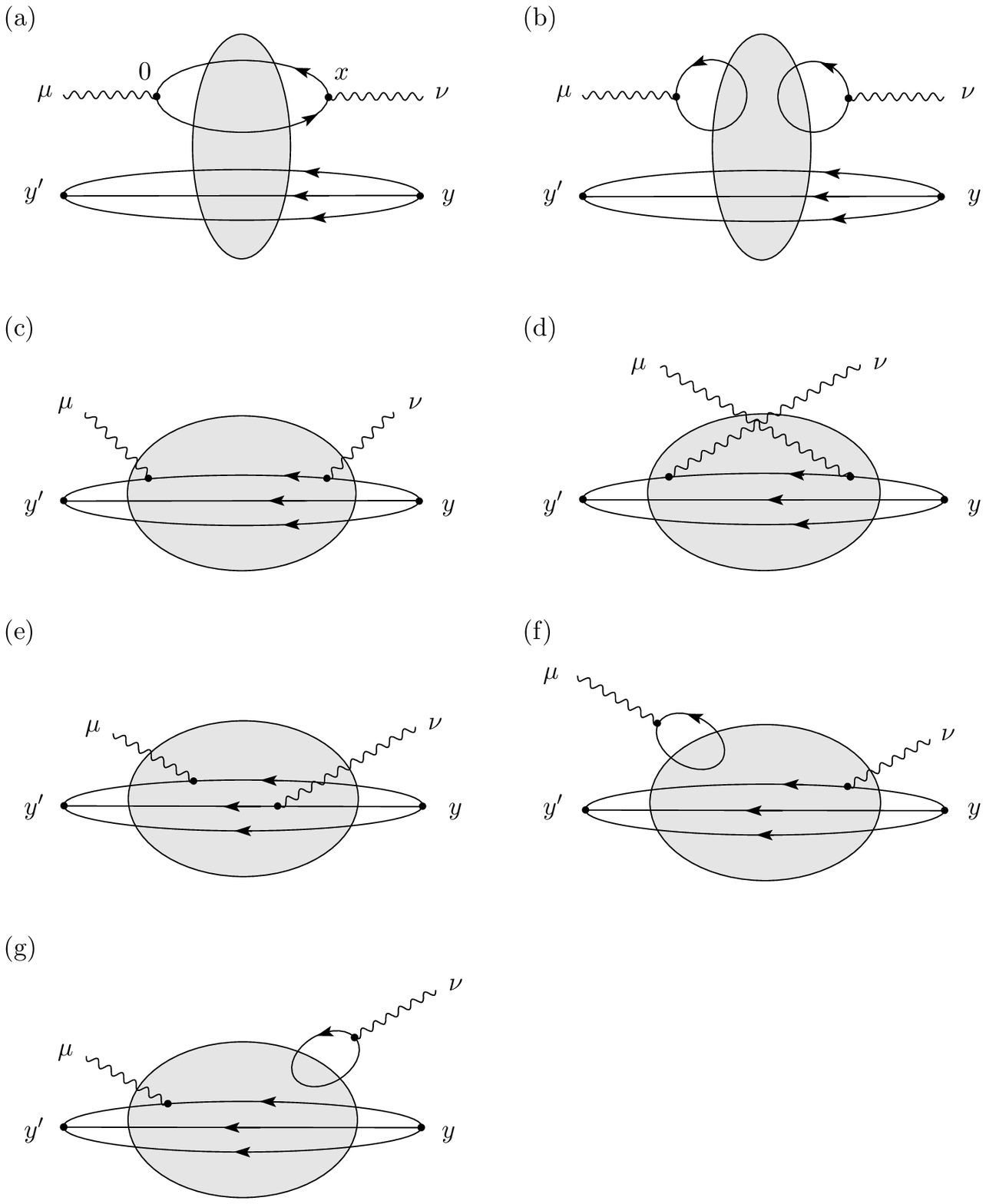}
\vspace*{.3cm}
\caption{Classification of contributions to the Compton amplitude 
according to their quark line skeleton
\label{fig2}}
\end{center}
\end{figure}

We consider the Compton amplitude for real or virtual photons, 
\begin{equation}
\label{2.1}
\gamma^{(*)}(q)+p(p)\,\,\,\longrightarrow\,\,\,
\gamma^{(*)}(q^\prime)+p(p^{\prime}) \,,
\end{equation}
and assume that $q^2=-Q^2\leq 0$ and $q^{\prime\,2}=-Q^{\prime\,2}\leq 0$. 
The matrix element for this process is 
\be
\label{2.3}
\mathcal{M}^{\mu\nu}_{s^{\prime}s}(p^{\prime},p,q)
=\frac{i}{2\pi m_p}\int d^4x~e^{-iqx}
\langle p(p^{\prime},s^{\prime})|\mbox{T}^*J^{\mu}(0)J^{\nu}(x)|p(p,s)\rangle
\,,
\ee
with the proton helicities $s,s'$ and with the electromagnetic current 
$J^{\lambda}(x)=\sum_q \bar{q}(x)Q_q\gamma^{\lambda}q(x)$, 
where $q=u,d,\dots$ are the quark field operators and $Q_q$ 
are the quark charges in units of the proton charge.
According to the LSZ formula we can write this matrix element as 
\begin{eqnarray}
\label{2.6}
\mathcal{M}^{\mu\nu}_{s^{\prime}s}(p^{\prime},p,q)&=&
-\frac{i}{2\pi m_pZ_p}
\int d^4y^{\prime}~d^4y~d^4x \, e^{ip^{\prime}y^{\prime}}
\bar{u}_{s^{\prime}}(p^{\prime})
(-i\rightslash_{y'}+m_p)
\nonumber
\\
&&{}
\left\langle\psi_p(y^{\prime})J^{\mu}(0)
e^{-iqx}J^{\nu}(x)\overline{\psi}_p(y)\right\rangle_{G,q,\bar{q}} \,
(i\leftslash_y+m_p)u_s(p)e^{-ipy}
\,,
\end{eqnarray}
where $\psi_p(x)=\Gamma_{\alpha\beta\gamma}
u_{\alpha}(x)u_{\beta}(x)d_{\gamma}(x)$ 
is an interpolating field operator for the proton. 
$\langle \dots \rangle_{G,q,\bar{q}}$ 
indicates a functional integral average over all gluon and quark field 
configurations with the weight $\exp[i \int d^4x \,{\cal L}_{\rm QCD}]$.  
Inserting the electromagnetic currents and performing 
the Gaussian integral over quark fields yields a decomposition of the 
amplitude as a sum over all possible contractions of quark field operators, 
that is a classification of contributions according to their quark line 
skeleton. Fig.\ \ref{fig2} shows the diagrammatic representation of 
these classes. 
The oriented lines represent nonperturbative quark propagators in 
the background of a gluon potential, indicated by the shaded blob, 
and we have to average over all possible gluon field configurations. 

A closer inspection of these diagram classes reveals that only the 
contributions from the classes (a) and (b) are leading at high energies, 
whereas the other classes correspond to fermion exchanges in the 
$t$-channel and are suppressed by powers of the energy. 
We will therefore not consider the latter any further. 
The perturbative expansion of the contributions of class (b) starts at 
higher order than that of class (a). At large photon virtualities 
class (b) is therefore suppressed with respect to class (a) by powers of 
$\alpha_s$. Moreover, 
it turns out that only class (a) gives rise to the usual dipole picture. 
Hence we will assume in the following: 
\begin{itemize}
\item[(i)]
The contribution of class (b) can be neglected. 
\end{itemize}
At low photon virtualities, however, the suppression of class (b) 
is lifted and there will be a correction to the dipole 
picture coming from this class. 

The contribution ${\cal M}^{(a)}$ to the Compton amplitude coming 
from the class (a) of Fig.\ (\ref{fig2}) can be expressed as 
\begin{eqnarray}\label{2.8}
\mathcal{M}^{(a)\mu\nu}_{s^{\prime}s}(p^{\prime},p,q)&=&
- \frac{i}{2\pi m_pZ_p}\int d^4y^{\prime}~d^4y \,
e^{ip^{\prime}y^{\prime}}\bar{u}_{s^{\prime}}(p^{\prime})(-i\rightslash_{y^{\prime}}
+m_p) 
\nonumber 
\\
&&{}
\sum_q Q^2_q  
\Big\langle
\wick{2}{<1\psi_p(y^{\prime})>1{\overline{\psi}}_p(y)}
A^{(q)\mu\nu}(q)\Big\rangle_G \,
(i\leftslash_y+m_p)u_s(p)e^{-ipy} \,,
\end{eqnarray}
where
\be
\label{2.8a}
A^{(q)\mu\nu}(q)=\int d^4x \,
\mathrm{Tr}\left[\gamma^{\mu}S^{(q)}_F(0,x;G)
e^{-iqx}\gamma^{\nu}S^{(q)}_F(x,0;G) \right] 
\,, 
\ee
and the $S_F^{(q)}$ are nonperturbative quark propagators in 
the gluon potential $G$. 
Here we insert next to the photon vertex (that is next to $\gamma^\nu$) 
factors of $1$ in the form 
\be
\label{2.11}
(i\rightslash_x-m_q)S_F^{(q,0)}(x,y)=-\delta^{(4)}(x-y)\,,
\ee
which contains a free quark propagator $S_F^{(q,0)}$ 
of (still arbitrary) mass $m_q$. 
Next we use the spin sum decompositions 
$\slash{k}+m_q = \sum_r u_r(k) \bar{u}(k)$ and 
$\slash{k}-m_q = \sum_r v_r(k) \bar{v}(k)$ in the spectral 
representation for the free quark propagator, 
\begin{eqnarray}\label{B.2}
S^{(q,0)}_F(x,y)&=&-\frac{1}{2\pi}\int^{\infty}_{-\infty}\frac{d\omega}{\omega+i\epsilon}
\int\frac{d^3k}{(2\pi)^32k^0}
\nonumber\\
&&{}
\sum_r\left\{e^{-ik_\omega x}u_r(k)\bar{u}_r(k)
e^{ik_\omega y}-e^{ik_\omega x}v_r(k)\bar{v}_r(k)e^{-ik_\omega y}\right\} 
\end{eqnarray}
with $k_\omega = (k^0+\omega, \mathbf{k})$, and insert this in 
$A^{(q)\mu\nu}$ of (\ref{2.8a}) and further in (\ref{2.8}) which 
finally leads us to an expression of $\mathcal{M}^{(a)\mu\nu}$ 
as a sum of four terms, 
\be
\label{Masumof4}
\mathcal{M}^{(a)\mu\nu}_{s^{\prime}s}(p^{\prime},p,q)=\sum^4_{j=1}
\mathcal{M}^{(a,j)\mu\nu}_{s^{\prime}s}(p^{\prime},p,q)\,,
\ee
the first of which reads 
\begin{eqnarray}\label{2.22}
\mathcal{M}^{(a,1)\mu\nu}_{s^{\prime}s}(p^{\prime},p,q)
\!&=&\!
\frac{1}{2\pi}\sum_qQ^2_q
\int\frac{d\omega}{\omega+i\epsilon}
\int\frac{d^3k}{(2\pi)^3 2k^0} \,
(q^0-k^0-k^{{\prime}0}-\omega+i\epsilon)^{-1} \,
(2k^{{\prime}0})^{-1}
\\
&&{}
\sum_{r^{\prime},r}
\langle \gamma(q^{\prime},\mu), p(p^{\prime},s^{\prime})
|\mathcal{T}^{(a)}|\bar{q}(k^{{\prime}}_\omega,r^{\prime}), 
q(k_\omega,r),p(p,s)\rangle \, 
\bar{u}_r(k)Z_q\gamma^{\nu}v_{r^{\prime}}(k^{{\prime}})\,,
\nonumber
\end{eqnarray}
where 
$k'=(\sqrt{m_q^2 + (\mathbf{q} - \mathbf{k})^2}, \mathbf{q} - \mathbf{k})$ 
and $k'_\omega=((q-k)^0 -\omega, \mathbf{q} - \mathbf{k})$. 
The other three terms contain the other combinations of spinors, 
$\bar{u} \gamma^\nu u$, $\bar{v} \gamma^\nu u$, and 
$\bar{v} \gamma^\nu v$. In those terms the spinors have different
momentum arguments, the matrix elements of ${\cal T}^{(a)}$ are crossed, 
and there appear different energy denominators. What we have 
done here is basically to cut open the trace in (\ref{2.8a}). Note that 
the quarks entering the matrix element are off the energy 
shell. The four terms have an interpretation which is reminiscent of 
old-fashioned perturbation theory~\cite{Ewerz:2004vf}, but we 
stress that our procedure has been completely nonperturbative so far. 
It turns out that only in the term in (\ref{2.22}) a quark and an 
antiquark come together with the proton in the incoming state of the 
matrix element, and thus only this term can give rise to the dipole 
picture. 

\begin{figure}
\begin{center}
\includegraphics[width=10cm]{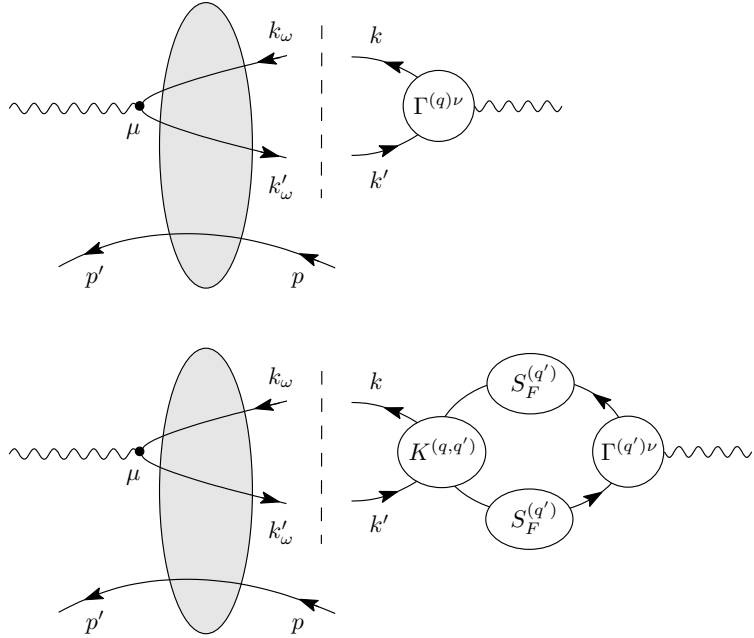}
\vspace*{.3cm}
\caption{The two terms for the amplitude $\mathcal{M}^{(a,1)}$ after 
applying (\ref{2.31}). 
The arrows on the lines indicate the interpretation as particle or antiparticle. 
The orientation of the momentum assignment to the lines is from 
right to left, irrespective of the arrows on the lines. 
\label{fig6}}
\end{center}
\end{figure}

In (\ref{2.22}) we have introduced the renormalisation factor $Z_q$ such 
that the ${\cal T}^{(a)}$-matrix element becomes a properly renormalised 
$T$-matrix element if quarks are assumed to have a mass shell. At the 
same time the factor $Z_q$ is necessary for the proper renormalisation of 
the photon-quark-antiquark vertex which has been widely discussed in the 
context of overlapping divergences in QED. It can be written here in terms 
of the renormalised vertex function $\Gamma^{(q)}$ and a rescattering 
term, 
\begin{equation}\label{2.31}
Z_q\gamma^{\nu}
=\Gamma^{(q)\nu}(l,l^{\prime})
 + \sum_{q'}\int K^{(q,q')}S^{(q')}_F \Gamma^{(q')\nu} S^{(q')}_F
\,.
\end{equation}
After inserting this back into (\ref{2.22}) we obtain two terms 
which are illustrated in Fig.\ \ref{fig6}. 
The dashed vertical line stands for the integration over $\omega$ and 
$d^3k$ in (\ref{2.22}). The quarks to the right of that line are on-shell, 
while those to the left are off the energy shell. 

The perturbative expansion of the rescattering term starts at higher order 
in $\alpha_s$ than the vertex term. At high $Q^2$ we can make the 
following assumption necessary for obtaining the dipole formula 
(\ref{dipolecross}): 
\begin{itemize}
\item[(ii)]
The rescattering term in (\ref{2.31}) is dropped and the vertex function 
$\Gamma^{(q) \nu}$ is taken at leading order, that is 
$\Gamma^{(q) \nu} = \gamma^\nu$. 
\end{itemize}
At lower photon virtualities we expect a correction to the dipole picture 
coming from the rescattering term. 

\section{High Energy Limit}
\label{sec:highenergy}

In the high energy limit, $|\mathbf{q}| \to \infty$, we find that 
$\Delta E =k^0+k^{\prime 0}-q^0 \, \to \, 0$. 
It turns out that of the four terms in (\ref{Masumof4}) 
only the term $\mathcal{M}^{(a,1)\mu\nu}$ 
exhibits a pinch singularity in the $\omega$-integration 
in this limit and hence gives the leading contribution 
which we can obtain in the form 
\begin{eqnarray}\label{3.14}
\mathcal{M}^{(a,1)\mu\nu}_{s^\prime s}(p^\prime, p,q)
&=& i \sum_qQ^2_q
\int \frac{d^3k}{(2\pi)^32k^02k^{\prime 0}} \,
(\Delta E)^{-1} \,
\theta\left(\bar{Q}^2-\tilde{m}^2(\alpha,\mathbf{k}_T)\right)
\\
&&{}
\sum_{r^\prime ,r} \, \langle \gamma(q^\prime,\mu),
p(p^\prime,s^\prime)|\mathcal{T}^{(a)}|\bar{q}(k^{\prime},r^\prime),
q(k,r),p(p,s) \rangle \, 
\bar{u}_r(k) \gamma^\nu v_{r^\prime}(k^\prime )\,.
\nonumber
\end{eqnarray}
We do not have enough space here to discuss in detail the $\theta$-function 
in this expression. We would only like to point out that it emerges 
naturally from the discussion of the pinch singularity and that it later on 
provides a regularisation of any singularity arising in the photon wave 
function of small distances. 

Contracting (\ref{3.14}) with the polarisation vectors for transversely 
or longitudinally polarised photons leads to the well-known expressions 
for the corresponding photon wave functions. Since (\ref{3.14}) is not 
separately gauge invariant, however, the longitudinal polarisation vector 
needs to be chosen such that its components remain finite in the high 
energy limit. Other choices lead to incorrect (and in fact arbitrary) 
results for the longitudinal photon wave function. 

Some further steps are required to finally obtain the dipole formula 
(\ref{dipolecross}). 
Obviously, we need to apply the procedure described above also to 
the outgoing photon. 
In order to be able to interpret the matrix elements in our formulae above 
(for example that of ${\cal T}^{(a)}$ in  (\ref{2.22})) as actual $T$-matrix 
elements we have to make the following assumption: 
\begin{itemize}
\item[(iii)]
Quarks have a mass shell $m_q$ and can be treated as asymptotic states. 
\end{itemize}
With this assumption we can then relate the $T$-matrix element to 
the reduced cross section and can be sure that the latter is non-negative. 
Assumption (iii) also allows us to define properly normalised dipole 
states which can then be interpreted as hadronic initial and final states. 

Finally, two further assumptions are needed in order to find the 
factorised form of Eq.\ (\ref{dipolecross}): 
\begin{itemize}
\item[(iv)]
At high energy the $T$-matrix for a dipole and a proton both in the incoming 
and in the outgoing state is diagonal in flavour, in the momentum fraction 
$\alpha$ carried by the quark as well as in the transverse separation 
$\mathbf{R}_T$ of the quark and antiquark in the dipole. 
\item[(v)]
The reduced matrix element $\sigma^{(q)}_{\rm red}$ depends only on 
$R_T^2$ and $s$, but is independent of the longitudinal momentum fraction 
$\alpha$. 
\end{itemize}
Clearly, there will be corrections to both of these assumptions from 
various sources. An example are the diagrams coming from the class (b) 
in Fig.\ (\ref{fig2}), which will become relevant at lower photon virtualities. 
It is quite obvious that these diagrams do in general not fulfil assumption (iv). 
Corrections are also expected from subleading diagrams in $\alpha_s$ 
in general. 

With the assumptions and approximations (i)-(v) above we finally 
obtain the dipole picture and the photon-proton cross section at 
high energy in the form (\ref{dipolecross}). 

\section{Summary}
\label{sec:summary}

Starting from a completely nonperturbative formulation of photon-proton 
scattering we have identified the assumptions and approximations 
(i)-(v) that 
are needed in order to obtain the dipole picture at high energies. 
At the same time we have found corrections to the dipole picture which 
can become large at small photon virtualities. We consider it an important 
task for the future to investigate in detail the validity of the assumptions, 
the accuracy of the approximations, and the size of the corrections. 
In our opinion these issues should be addressed in order to put the 
results obtained in the framework of the dipole picture on solid ground. 

The framework developed here should be suitable for studying 
the effects caused by the non-existence of a mass-shell for quarks, and 
for using nonperturbative quark propagators, obtained for example 
from Dyson-Schwinger equations or from lattice simulations, in 
phenomenology. 

\section*{Acknowledgments}
C.\,E.\ was supported by a Feodor Lynen fellowship of the
Alexander von Humboldt Foundation.

\end{document}